\input nature_pp.sty
\input psfig

\centerline{\bfb Early Planet Formation as a Trigger for further Planet
            Formation$^1$}

\medskip
\centerline{\bfb Philip~J.~Armitage$^2$ \& Brad~M.S.~Hansen$^3$ }
\smallskip    
Canadian Institute for Theoretical Astrophysics, University of Toronto,
Toronto, ON, M5S 3H8, Canada

\bigskip

{\ssbig Recent discoveries of extrasolar  
planets{\refto{MUQ98}$^,$\refto{MB98}} at 
small orbital radii, or with significant eccentricities, indicate that
interactions between massive planets and the disks of gas and dust
from which they formed are vital for determining the final shape 
of planetary 
systems{\refto{LBR96}{$^,$}\refto{MHHT98}{$^,$}\refto{Rasio}{$^,$}\refto{WM96}}. 
We show that if this interaction occurs at an early epoch, when the 
protoplanetary disc was still massive, then rapid planet growth 
through accretion causes an otherwise stable disc to fragment 
into additional planetary mass bodies when the planetary mass 
reaches $4-5 m_{\rm Jupiter}$. We suggest that 
such catastrophic planet formation could account for apparent differences 
in the mass function of massive planets and brown dwarfs\refto{MUQ98}, and 
the existence of young stars that appear to have dissipated their discs 
at an early epoch\refto{Strom95}. Subsequent gravitational 
interactions{\refto{Rasio}{$^,$}\refto{WM96}{$^,$}\refto{Brett}{$^,$}\refto{LI97}} 
will lead to planetary systems comprising a small number of 
massive planets in eccentric orbits.}
\medskip

\footnote{}{$^1$To appear in Nature, 9th December 1999.}

\footnote{}{$^2$Present address: 
             Max-Planck-Institut for Astrophysik,
             Karl-Schwarzschild-Str. 1,
             D-85740 Garching, Germany}
	     
\footnote{}{$^2$Present address: 
             Department of Astrophysical Sciences,
             Princeton University, 
             Peyton Hall, Ivy Lane, 
             Princeton, NJ 08544-1001}
	     
The planet--disc interaction has been studied extensively 
for low mass protoplanetary discs{\refto{LBR96}{$^,$}\refto{AL96}{$^,$}\refto{TML96}{$^,$}
\refto{LP79}{$^,$}\refto{K99}{$^,$}\refto{Bryden99}}.
This is appropriate for the proto-Solar nebula where the rate limiting 
step, the assembly of the cores of the giant planets from smaller 
bodies\refto{S69}, is believed to require timescales comparable to 
the lifetimes of protoplanetary 
discs, which are observed\refto{Strom95} to last for a 
${\rm few} \times 10^{6} - 10^{7} \ {\rm yr}$. However, some extrasolar 
giant planets could form more rapidly, either via direct hydrodynamic 
collapse{\refto{Boss}}, 
or via accelerated core formation in discs that are significantly 
more massive\refto{OB95} than the minimum 
mass Solar nebula\refto{HNN85}. More broadly, if 
angular momentum transport mechanisms other than self-gravity are inefficient in 
discs where the ionization fraction is low (and no purely hydrodynamic 
instabilities that lead to outward angular momentum transport are known 
to exist in Keplerian disc flows\refto{BHS96}), then the outer regions 
of protoplanetary discs may remain only marginally stable against gravitational 
instability even at {\it late} evolutionary epochs{\refto{Larson}{$^,$}\refto{LP90}}.
In either case, planet-disc interactions could occur while the effects of disc 
self-gravity are still important.

The local stability of a gaseous disc against gravitational instability depends 
upon the balance between thermal pressure and self-gravity. At a point in a disc 
with sound speed $c_s$, surface density $\Sigma$, and angular velocity $\Omega$, 
the controlling parameter\refto{Toomre} is Toomre's $Q$, defined as,
$$ 
 Q = { { c_s \Omega } \over {\pi G \Sigma} },
$$
with $G$ the gravitational constant. Numerical 
simulations{\refto{LB94}{$^,$}\refto{NBAA98}} 
show that non-axisymmetric instabilities set in at $Q \ltorder 1.5$, and become 
increasingly violent for smaller values of $Q \approx 1$. We consider the 
relatively cool outer regions of the disc, at radii of several a.u., and
set our initial conditions such that the disc is both marginally unstable, and 
has properties comparable to the upper end of the mass distribution 
of T Tauri discs inferred from mm wavelength observations\refto{OB95},
which have $m_{\rm disc} \approx 0.1 \ m_\odot$. This is around an 
order of magnitude greater than the canonical minimum mass Solar nebula 
value of $\sim 10^{-2} \ m_\odot$, though even for our Solar 
System the initial disc mass may have been substantially in excess of this 
minimum{\refto{L87}$^,$\refto{P96}}. The surface density profile 
is taken to be
$$ 
 \Sigma = \Sigma_0 r^{-3/2} \left( 1 - \sqrt{r_{\rm in} \over r} \right), 
$$
where $r_{\rm in}$ is the inner edge of the simulated disc annulus. 
$\Sigma_0$ and $c_s$ are chosen such 
that $m_{\rm disc} = 0.1$, and the 
ratio of disc scale height to radius at the outer edge is $(h/r) = 0.075$. 
With these parameters $Q > 1.5$ at all radii, and the  
disc is everywhere close to stable against gravitational instability, as 
expected if it is the endpoint of an earlier phase of violent gravitational 
instabilities that drive rapid angular momentum transport\refto{LP90}.

Figure 1 shows the evolution of the disc, computed using a Lagrangian 
hydrodynamics code. The isolated disc is shown at $t=512$, in  
units where $\Delta t = 1$ corresponds to the dynamical time, 
$\Omega^{-1}$, at $r_{\rm in}$. This run shows 
weak spiral arms in the outer disc, as expected from the $Q$ profile 
on the basis of previous simulations of gravitationally unstable 
discs{\refto{LB94}$^,$\refto{NBAA98}}, and is amply stable against 
fragmentation. The disc 
surface density does not evolve significantly over this relatively 
short interval, as expected for a thin disc where 
the efficiency of angular momentum transport from gravitationally 
instabilities, if parameterized approximately via an equivalent 
Shakura-Sunyaev $\alpha$ parameter\refto{SS}, corresponds to a 
fairly small effective $\alpha \sim 10^{-2}$.

We now consider the evolution of the same star-disc system with an
embedded planet of initial mass $m_p = 10^{-3} m_*$. 
For seeds of this mass and smaller, the additional 
potential fluctuations induced 
by the planet at the Lindblad resonances\refto{LP79} are 
small compared to the  
background fluctuations due to the disc's own self-gravity, 
as measured in the control run. This is shown in Figure 2. Neither the mass resolution 
nor the equation of state are realistic enough to model the internal structure of
individual planets, so we focus solely on their influence on the disc, 
for which purpose details of their internal structure are unimportant.

The presence of a Jupiter mass planet significantly modifies the
disc evolution. A partial gap is cleared in the disc on the dynamical 
timescale at $r_p$, bounded by a strongly compressed gravitational 
wake attached to the planet. This forms part of an $m=2$ pattern 
of strong spiral arms, along with weaker transient spiral features 
excited in the disc by the combination of gravitational instability 
and planetary perturbation. The presence of a gap fails to prevent 
ongoing accretion along the spiral arms at a rate $\dot{m}_p \propto m_p$, with an e-folding 
time of a few planetary orbits. Continuing accretion is expected since 
the disc viscosity needs to be significantly lower than the values 
expected in a gravitationally unstable disc to inhibit accretion 
altogether\refto{TML96}. Much of this mass accumulates in a resolved, 
strongly tidally distorted disc surrounding the planet. As the 
planet mass grows, the overdensity in the spiral arms and at the 
Lindblad resonances increases while 
the background surface density profile is unable to evolve on 
as rapid a timescale. This essential imbalance in timescales is 
expected to be valid even for the formation of 
giant planets in a minimum mass Solar nebula\refto{P96}, and is 
therefore a robust prediction for the more massive discs
studied here. The rapid growth in mass leads to an increased amplitude 
of potential fluctuations, as shown in Figure 2,  
decreasing disc stability, and inevitable fragmentation 
at the gap edges, shown in the lower panels of Figure 1. For these 
disc parameters and 
equation of state this occurs at $m_p = 4-5 \times 10^{-3} m_*$. Once this mass
is reached, rapid fragmentation into numerous planetary mass 
bodies occurs near both the inner and 
outer Lindblad resonances. 

Several additional calculations were used to check the sensitivity 
of the results to the initial conditions and numerical method. With 
the same initial conditions, fragmentation occurs at the same final planet mass 
in lower resolution simulations with 10,000 and 20,000 particles, 
though the initial masses of fragments do vary. In all 
cases, however, the fragments accrete rapidly from the disc, and 
so their final masses would fall into the regime of massive planets.
Fragmentation also occurs at the same planet mass in a simulation 
where the initial seed mass was $m_p = 2 \times 10^{-4} m_*$, which 
is closer to the mass of a giant planet core beginning runaway 
accretion of disc gas\refto{P96}. For this simulation
the time required prior to fragmentation was roughly doubled. Fragmentation 
does {\it not} occur if we artificially set the mass accreted by a   
Jupiter mass seed to zero, verifying that it is the increased planetary mass after 
significant accretion that leads to instability. 

The existence of a propagating mode of planet formation has 
implications for the evolution of protoplanetary discs and the statistics 
of planetary systems. In particular, the formation of one massive 
planet could suffice to trigger rapid planet formation across the 
range of disc radii for which $Q \approx 1-2$. This could allow 
massive planets to form at large radii where 
the timescales for planet formation via other mechanisms 
can become worryingly long compared to typical disc lifetimes.
The consequent disruption of the outer disc concomitant with such 
violent planet formation would allow the unreplenished inner disc to drain 
viscously onto the star in a short timescale. Studies of the UV and H$\alpha$ 
flux arising from the accretion process, and near infra-red flux from the inner 
disc, suggest that a significant fraction of T Tauri 
stars are able to dissipate their inner discs rapidly\refto{Strom95}.
Related processes may be relevant
to the formation of planetary satellite systems\refto{LP79}.

For planetary formation, these results imply that steady 
growth of giant planets in massive discs around solar mass stars 
is limited by the vulnerability of the disc to fragmentation once the 
planetary mass reaches approximately $5 m_{\rm Jupiter}$. The resulting formation of 
additional planets, which then compete to accrete the available disc gas, 
implies an upper limit to the mass of massive planets formed via this 
mechanism. In particular, 
even if the disc was sufficiently massive it would not be possible 
to grow a planet from a Jupiter mass far into the brown dwarf regime. 
This is consistent with observational evidence that planets and brown 
dwarfs do not share a common mass function\refto{MUQ98}, which has prompted 
suggestions that a break in formation mechanisms exists at 
around $7 m_{\rm Jupiter}$. Finally we note that the endpoint of 
early disc fragmentation would be a system of numerous massive coplanar 
planets in initially close to circular orbits. Such a system 
would possess a global organisation imprinted via non-local 
gravitational effects at birth. The  
subsequent evolution will be strongly affected by mutual perturbations.
These would be favourable initial conditions for the eventual 
formation of a system comprising one or more massive planets on 
eccentric orbits\refto{LI97}. 

\references
{\parskip=0pt

\refis{Rasio} Rasio, F. A., \& Ford, E. B., 
Dynamical instabilities and the formation of extrasolar planetary systems, 
{\it Science}, {\bf 274}, 954-965 (1996) \medskip \par

\refis{WM96} Weidenschilling, S. J., \& Marzari, F., 
Gravitational scattering as a possible origin for giant
planets at small stellar distances, 
{\it Nature}, {\bf 384}, 619-621 (1996) \medskip \par

\refis{MB98} Marcy, G. W., \& Butler, R. P., 
Detection of extrasolar giant planets, 
{\it Ann. Rev. Astron. Astrophys.}, {\bf 36}, 57-98 (1998) \medskip \par

\refis{LBR96} Lin, D. N. C., Bodenheimer, P., \& Richardson, D. C., 
Orbital migration of the planetary companion of 51 Pegasi to its
present location, 
{\it Nature}, {\bf 380}, 606-607 (1996) \medskip \par

\refis{MHHT98} Murray, N., Hansen, B., Holman, M., \& Tremaine, S., 
Migrating planets, 
{\it Science}, {\bf 279}, 69 (1998) \medskip \par

\refis{AL96} Artymowicz, P., \& Lubow, S. H., 
Mass flow through gaps in circumbinary disks, 
{\it Astrophys. J.}, {\bf 467}, L77-L80 (1996) \medskip \par

\refis{TML96} Takeuchi, T., Miyama, S. M., \& Lin, D. N. C., 
Gap formation in protoplanetary disks, 
{\it Astrophys. J.}, {\bf 460}, 832-847 (1996) \medskip \par

\refis{LP79} Lin, D. N. C., \& Papaloizou, J., 
On the structure of circumbinary accretion discs and 
the tidal evolution of commensurable satellites, 
{\it Mon.\ Not.\ R.\ Astron.\ Soc.}, {\bf 188}, 191-201 (1979)
\medskip \par

\refis{K99} Kley, W., 
Mass flow and accretion through gaps in accretion discs,
{\it Mon.\ Not.\ R.\ Astron.\ Soc.}, {\bf 303}, 696 (1999) \medskip \par

\refis{Bryden99} Bryden, G., Chen, X., Lin, D. N. C., Nelson, R. P., \& 
Papaloizou, J. C. B., 
Tidally induced gap formation in protostellar disks: Gap 
clearing and suppression of protoplanetary growth, 
{\it Astrophys. J.}, {\bf 514}, 344-367 (1999) \medskip \par

\refis{Strom95} Strom, S. E., 
Initial frequency, lifetime and evolution of YSO disks, 
{\it Rev. Mex. Astron. Astrophys. Conf. Ser.}, {\bf 1}, 317-328 (1995)
\medskip \par

\refis{Larson} Larson, R. B., 
Gravitational torques and star formation, 
{\it Mon.\ Not.\ R.\ Astron.\ Soc.}, {\bf 206}, 197-207 (1984)
\medskip \par

\refis{LP90} Lin, D. N. C., \& Pringle, J. E., 
The formation and initial evolution of protostellar disks, 
{\it Astrophys. J.}, {\bf 358}, 515-524 (1990) \medskip \par

\refis{BHS96} Balbus, S. A., Hawley, J. F., \& Stone, J. M., 
Nonlinear stability, hydrodynamical turbulence, and transport in disks, 
{\it Astrophys. J.}, {\bf 467}, 76-86 (1996) \medskip \par

\refis{Benz} Benz, W., 
Smooth particle hydrodynamics -- A review, 
in {\it The numerical modelling of nonlinear stellar pulsations}, 
ed J. R. Buchler, Kluwer Academic Publishers (Dordrecht), 269-287
(1990) \medskip \par

\refis{Monaghan} Monaghan, J. J., 
Smoothed particle hydrodynamics, 
{\it Ann. Rev. Astron. Astrophys.}, {\bf 30}, 543-574 (1992) \medskip \par

\refis{BH86} Barnes, J., \& Hut, P., 
A hierarchical O(N log N) force calculation algorithm, 
{\it Nature}, {\bf 324}, 446 (1986) \medskip \par

\refis{Toomre} Toomre, A., 
On the gravitational stability of a disk of stars, 
{\it Astrophys. J.}, {\bf 139}, 1217-1238 (1964) \medskip \par

\refis{LB94} Laughlin, G., \& Bodenheimer, P., 
Nonaxisymmetric evolution in protostellar disks, 
{\it Astrophys. J.}, {\bf 436}, 335-354 (1994) \medskip \par

\refis{NBAA98} Nelson, A. F., Benz, W., Adams, F. C., \& Arnett, D., 
Dynamics of circumstellar disks, 
{\it Astrophys. J.}, {\bf 502}, 342-371 (1998) \medskip \par

\refis{OB95} Osterloh, M., \& Beckwith, S. V. W., 
Millimeter-wave continuum measurements of young stars, 
{\it Astrophys. J.}, {\bf 439}, 288-302 (1995) \medskip \par

\refis{Boss} Boss, A. P., 
Evolution of the solar nebula IV. Giant gaseous protoplanet formation, 
{\it Astrophys. J.}, {\bf 503}, 923-937 (1998) \medskip \par

\refis{SS} Shakura, N. I., \& Sunyaev, R. A., 
Black holes in binary systems. Observational appearance, 
{\it Astron. Astrophys.}, {\bf 24}, 337-355 (1973) \medskip \par

\refis{MUQ98} Mayor, M., Udry, S., \& Queloz, D., 
The mass function below the substellar limit, 
in {\it The Tenth Cambridge Workshop on Cool Stars,
Stellar Systems and the Sun}, ASP Conf. Ser. 154, eds R. A. Donahue \& 
J. A. Bookbinder, 77-87 (1998) \medskip \par

\refis{HNN85} Hayashi, C., Nakazawa, K., \& Nakagawa, Y., 
Formation of the solar system, 
in {\it Protostars and planets II}, eds D. C. Black \& M. S. Matthews, 
Univ. of Arizona Press (Tucson), p.~1100-1153 (1985) \medskip \par

\refis{LI97} Lin, D. N. C., \& Ida, S., 
On the origin of massive eccentric planets, 
{\it Astrophys. J.}, {\bf 477}, 781-791 (1997) \medskip \par

\refis{L87} Lissauer, J. J., 
Timescales for planetary accretion and the structure of the 
protoplanetary disk, 
{\it Icarus}, {\bf 69}, 249-265 (1987) \medskip \par

\refis{P96} Pollack, J. B., Hubickyj, O., Bodenheimer, P., Lissauer, J. J., 
Podolak, M., \& Greenzweig, Y., 
Formation of the giant planets by concurrent accretion of solids and gas, 
{\it Icarus}, {\bf 124}, 62-85 (1996) \medskip \par

\refis{S69} Safronov, V. S., 
Evolution of the protoplanetary cloud and formation of the Earth 
and the planets, 
Nauka (Moscow), (1969), 
(English translation for NASA and NSF by Israel Program for 
Scientific Translations, NASA-TT-F-677, 1972) \medskip \par

\refis{Brett} Gladman, B., 
Dynamics of systems of two close planets, 
{\it Icarus}, {\bf 106}, 247-263 (1993) \medskip \par

}
\endreferences
\endmode
\bigskip
\par\noindent ACKNOWLEDGEMENTS. We thank Norm Murray for many helpful 
discussions, and Norman Wilson for maintaining the required computational 
resources.

\bigskip

Correspondence to P. Armitage (email: armitage@mpa-garching.mpg.de)

\vfill
\eject

\psfig{figure=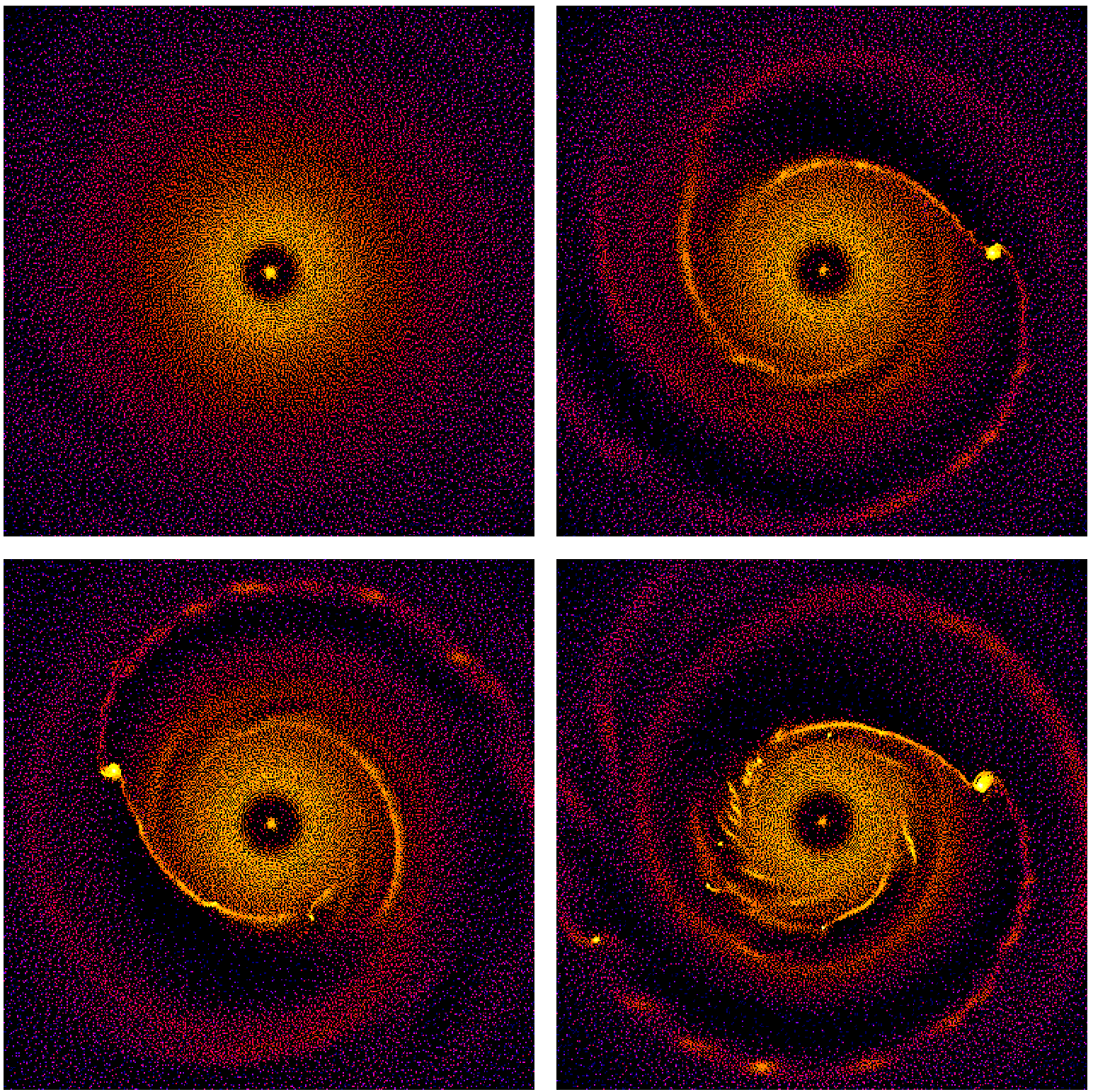,width=6.0in,height=6.0in}

\bigskip

\noindent{\bf Figure~1 -- Disc surface density:} Disc structure, 
computed using a Smooth Particle Hydrodynamics (SPH)
code{\refto{Benz}{$^,$}\refto{Monaghan}}, with individual timesteps 
for the particles and a tree structure for computing the gravitational 
forces{\refto{BH86}}. We use 60,000 SPH particles, an isothermal 
equation of state, and standard artificial viscosity
parameters\refto{Monaghan}. We have 
verified that torques due to gravity and pressure forces dominate 
over those attributed to artificial viscosity in determining the 
planet-disc evolution. Additional support for collapsed 
objects is provided by limiting the force resolution via a minimum 
SPH smoothing length, $h_{\rm min} = 0.075$. The central star is 
treated as a smoothed point mass chosen to give a Keplerian potential 
at $r > r_{\rm in}$. We use units in which $r_{\rm in} = m_* = 1$, and
set the outer disc edge at $r_{\rm out} = 25$. The upper
left panel shows the isolated disc evolved to a time $t=512$, the upper right 
panel the disc at the same time but with a planet, initially of $10^{-3} \ m_*$, 
orbiting in a coplanar circular orbit at $r_p = 12.5$. The planet is
treated as a point mass smoothed on a scale $h_p = 0.1$. The colours 
signify density on a logarithmic scale.
In the presence of a planet the 
spiral structure is significantly amplified, and a partial gap has been 
cleared in the disc material. By $t=608$ (lower left), when the planetary mass (plus 
surrounding circumplanetary disc) has reached $4-5 \times 10^{-3} \ m_*$,
the disc near the inner Lindblad resonance has become unstable and 
fragments into additional planets. Thereafter rapid destruction of the 
disc occurs. By $t=736$ (lower right) numerous fragments have formed, including 
one near the outer Lindblad resonance, and are accreting rapidly.

\vfill
\eject

\psfig{figure=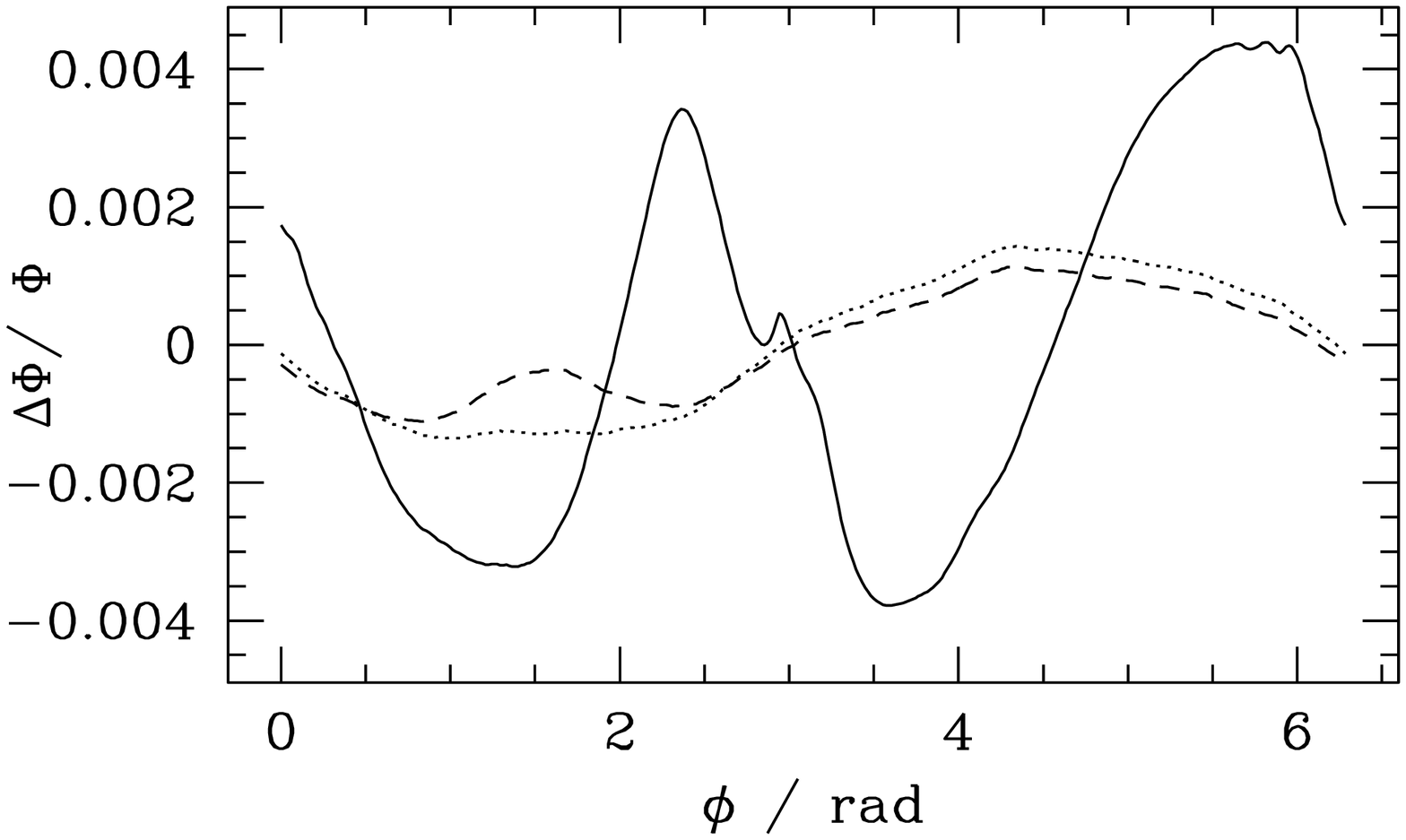,width=4.0truein,height=4.0truein}

\bigskip

\noindent{\bf Figure~2 -- Gravitational potential fluctuations}, evaluated 
as a function of azimuthal angle $\phi$ at the inner Lindblad resonance where 
fragmentation first occurs. The amplitude of potential fluctuations in the 
control run (dotted line) is not significantly increased by the addition of 
a Jupiter mass seed planet (dashed line). By $t=592$ (solid line), the 
greatly increased planet mass has led to a strong $m=2$ mode. Fragmentation 
occurs shortly afterwards.

\vfill\eject

\bye